\documentclass[aps,showpacs,twocolumn,floats,superscriptaddress]{revtex4}
\usepackage{hyperref}
\usepackage{graphicx}
\usepackage{color}

\usepackage{amsmath}

\begin{document}
\title{Magnetoconductance signatures of subband structure in semiconductor nanowires}

\author{Gregory W. Holloway}
\affiliation{Institute for Quantum Computing, University of Waterloo, Waterloo, Ontario, N2L 3G1, Canada}
\affiliation{Department of Physics and Astronomy, University of Waterloo, Waterloo, Ontario, N2L 3G1, Canada} 
\affiliation{Waterloo Institute for Nanotechnology, University of Waterloo, Waterloo, Ontario N2L 3G1, Canada}

\author{Daryoush Shiri}
\affiliation{Institute for Quantum Computing, University of Waterloo, Waterloo, Ontario, N2L 3G1, Canada}

\author{Chris M. Haapamaki}
\affiliation{Institute for Quantum Computing, University of Waterloo, Waterloo, Ontario, N2L 3G1, Canada}
\affiliation{Department of Chemistry, University of Waterloo, Waterloo, Ontario, N2L 3G1, Canada}
\affiliation{Department of Engineering Physics, Centre for Emerging Device Technologies, McMaster University, Hamilton, Ontario L8S 4L7, Canada}

\author{Kyle Willick}
\affiliation{Institute for Quantum Computing, University of Waterloo, Waterloo, Ontario, N2L 3G1, Canada} 
\affiliation{Department of Physics and Astronomy, University of Waterloo, Waterloo, Ontario, N2L 3G1, Canada}
\affiliation{Waterloo Institute for Nanotechnology, University of Waterloo, Waterloo, Ontario N2L 3G1, Canada}

\author{Grant Watson}
\affiliation{Institute for Quantum Computing, University of Waterloo, Waterloo, Ontario, N2L 3G1, Canada} 

\author{Ray R. LaPierre}
\affiliation{Department of Engineering Physics, Centre for Emerging Device Technologies, McMaster University, Hamilton, Ontario L8S 4L7, Canada}

\author{Jonathan Baugh\footnote{corresponding author: baugh@iqc.ca}}
\affiliation{Institute for Quantum Computing, University of Waterloo, Waterloo, Ontario, N2L 3G1, Canada}
\affiliation{Department of Chemistry, University of Waterloo, Waterloo, Ontario, N2L 3G1, Canada}
\affiliation{Waterloo Institute for Nanotechnology, University of Waterloo, Waterloo, Ontario N2L 3G1, Canada}

\begin{abstract}
The radial confining potential in a semiconductor nanowire plays a key role in determining its quantum transport properties. Previous reports have shown that an axial magnetic field induces flux-periodic conductance oscillations when the electronic states are confined to a shell.  This effect is due to the coupling of orbital angular momentum to the magnetic flux. Here, we perform calculations of the energy level structure, and consequently the conductance, for more general cases ranging from a flat potential to strong surface band bending. The transverse states are not confined to a shell, but are distributed across the nanowire. It is found that, in general, the subband energy spectrum is aperiodic as a function of both gate voltage and magnetic field. In principle, this allows for precise identification of the occupied subbands from the magnetoconductance patterns of quasi-ballistic devices. The aperiodicity becomes more apparent as the potential flattens. A quantitative method is introduced for matching features in the conductance data to the subband structure resulting from a particular radial potential, where  a functional form for the potential is used that depends on two free parameters. Finally, a short-channel InAs nanowire FET device is measured at low temperature in search of conductance features that reveal the subband structure. Features are identified and shown to be consistent with three specific subbands. The experiment is analyzed in the context of the weak localization regime, however, we find that the subband effects predicted for ballistic transport should remain visible when back scattering dominates over interband scattering, as is expected for this device. 
\end{abstract}

\maketitle

\section{Introduction}
\indent The study of quantum transport of electrons and holes in semiconductor nanowires is of fundamental interest, and underlies recent developments in nanoscale sensing \cite{du2009, Salfi2010} and potential avenues for quantum information processing \cite{Flindt07, schroer2011, Nadj-Perge10, Sau2010, Mourik2012}. The quasi one-dimensional (1D) geometry of nanowires allows for a wide range of experiments on low dimensional transport, but correct interpretation of results often requires a detailed understanding of the transverse subband structure due to the confining radial electrostatic potential. Precise knowledge of the radial potential, however, is not usually straightforward to determine experimentally. Several recent experiments have shed light on the subband structure in multi-band nanowires. Quantized conductance steps were observed in quasi-ballistic (short channel) InAs nanowire field-effect transistors (FETs) \cite{Ford2012,Chuang2013}, and attributed to the successive occupation of the first few subbands. In the presence of a perpendicular magnetic field, these steps split into two due to the Zeeman interaction. The resulting conductance patterns have been observed as a function of magnetic field and gate voltage \cite{Vigneau2014, VanWeperen2013}. The presence of an axial field produces qualitatively different conductance patterns due to the coupling of orbital angular momentum to magnetic flux. Axial field magnetoconductance studies of InN nanowires \cite{Richter2008,Blomers2008} and InAs nanowires\cite{Blomers2011} reveal oscillations caused by the occupation of orbital angular momentum subbands. With strong surface band bending, a cylindrical conducting shell forms below the nanowire surface and the resulting conduction electron energy levels are parabolic in magnetic field \cite{Blomers2008}. Levels with adjacent angular momentum quantum numbers are shifted from each other by one flux quantum. This gives rise to a flux-periodic, diamond shaped energy level structure, so that varying magnetic field at a fixed chemical potential leads to flux-periodic conductance oscillations as the occupation of orbital states is modulated. These flux-periodic oscillations have been observed in InN nanowires \cite{Richter2008,Blomers2008}, however the precise orbital states contributing to conductance were not identified. Experiments on GaAs/InAs core-shell nanowires \cite{Rosdahl2014,Demarina2014}, where conductance is predominantly due to the shell, also showed flux periodic oscillations. Importantly, the phase of the oscillations was seen to change by $\pi$ at certain gate voltages, as would be expected from the diamond-shaped pattern of orbital energy levels. In all of these axial field magnetoconductance experiments, the focus has been on conduction in a thin shell close to the nanowire surface, such that flux-periodic oscillations are expected. This is not the general case, as different materials and surface conditions can give rise to varying degrees of surface band bending. For example, nanowires with an epitaxial larger bandgap shell are expected to have reduced band bending \cite{Holloway2013a,Tilburg2010}, giving more uniformly distributed transverse electronic wavefunctions. Bare InAs nanowires have not previously shown the expected flux-periodic oscillations\cite{Blomers2011}, perhaps due to reduced surface band bending compared to InN nanowires. These examples reflect the need to model transverse subbands for more general radial potentials to accurately model electronic transport. In this paper, it is found that lower surface confinement alters the shape of the transverse subband energy spectrum and its dependence on magnetic field to have a lower degree of periodicity, making precise identification of orbital subbands and estimation of the radial potential a practical possibility. \\
\indent Here, we calculate the energy spectra of transverse subbands for various radial potentials, ranging from flat to those with strong surface band bending. We find a quasi-parabolic behavior of these energies with respect to magnetic field, but with large variations in curvature depending on the radial potential and on the radial quantum number. Indeed, the energetic ordering of the subbands depends on the degree of band bending, and the overall pattern of conductance versus magnetic field and gate voltage provides a fingerprint of the underlying radial potential. Although similar studies have been applied to InN and GaAs/InAs core-shell nanowires, the wavefunctions in those cases are assumed to be confined in a thin conducting shell, either by the core-shell structure or by a strong surface potential. Here, we consider the more general case of a wavefunction that extends across the nanowire cross-section, enabling the description of devices over a wide range of surface potentials. In addition, we report the results of low temperature conductance measurements on a short-channel InAs nanowire FET as a function of gate voltage and magnetic field. Features are identified in the magnetoconductance data that are quantitatively consistent with a particular assignment of states and radial potential, although the quality of the data falls short of an unambiguous assignment. By calculating conductance in the weakly localized regime, we find a consistent description of its general magnitude over a gate voltage range of 6 volts. We present a method of analyzing magnetoconductance data to find a matching radial potential, and suggest that it is best applied to nanowire devices in the ballistic or quasi-ballistic regime.   

\begin{figure}[t]
\includegraphics[width= 8.6cm]{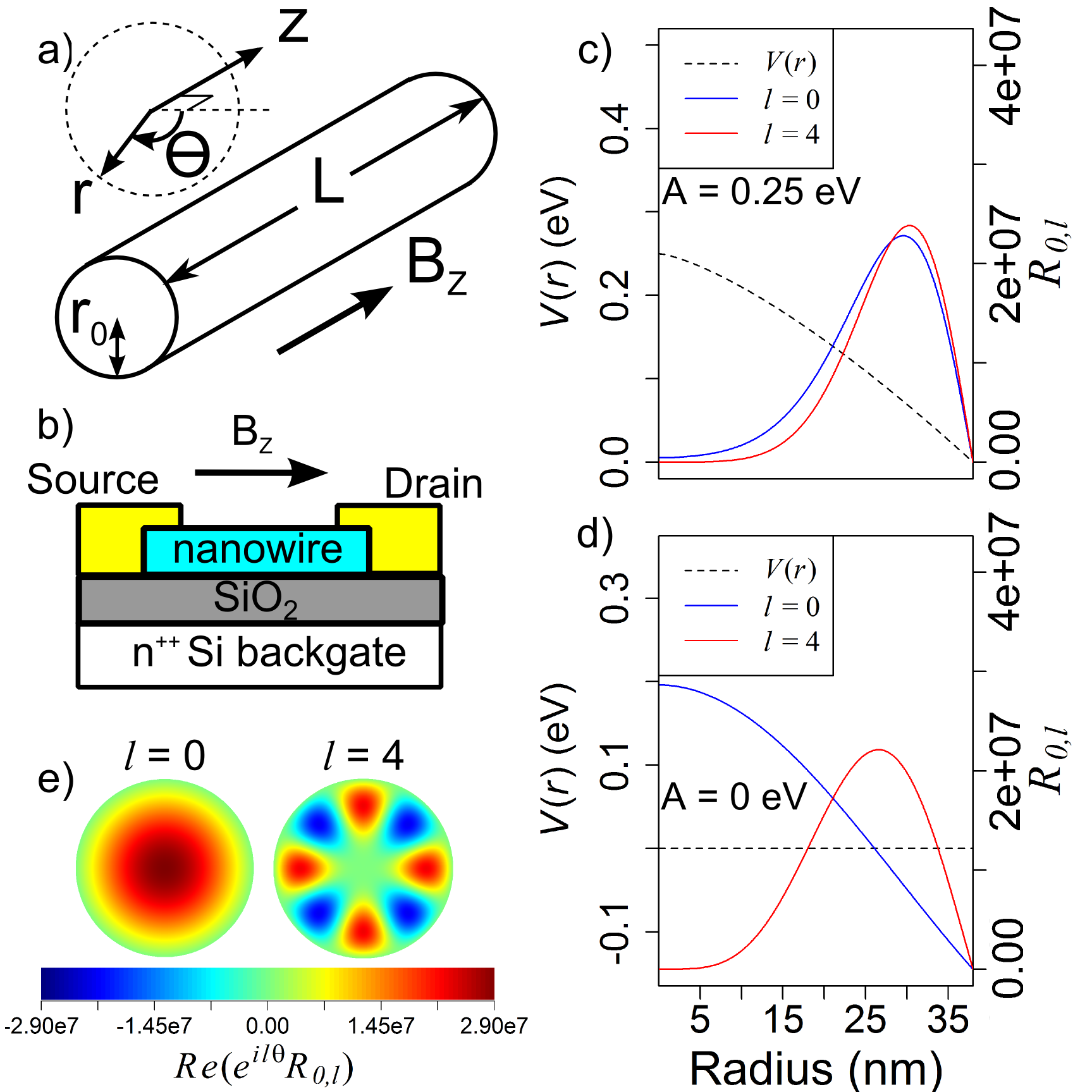} 
\caption{(a) The cylindrical nanowire geometry is shown with an axial magnetic field $B_z$. (b) Schematic of the nanowire FET used to measure magnetoconductance. The two-terminal conductance is measured between the source and drain contacts as a function of $B_z$ and gate voltage. (c,d) Radial wavefunctions $R_{0,0}(r)$ and $R_{0,4}(r)$, normalized by $\int_0^\infty\int_0^{2\pi}|e^{il\theta}R_{n,l}(r)|^2rdrd\theta = 1$, calculated for a cylindrically symmetric radial potential $V(r)$ defined in the main text with $b$ = 2.75 and (c) $A$ = 0.25 eV, (d) $A$ = 0 eV. $R_{n,l}(r)$ is characterized by the radial and angular quantum numbers $n$ and $l$, respectively. The effective mass used is for InAs. Strong band bending results in a wavefunction proximate to the nanowire surface for all states. (e) Real part of the transverse electron wavefunction for the two states shown in (d).} \label{fig1}
\end{figure}

\section{Model}
\indent Consider a nanowire of radius $r_0$ and length $L>2r_0$, as shown schematically in figure 1a. Assuming cylindrical symmetry, the single particle wavefunction for conduction electrons can be written as the product: $\psi(r,\theta,z) = e^{ikz}e^{il\theta}R_{n,l}(r)$, where $(r,\theta,z)$ are cylindrical coordinates, $k$ is the axial wavenumber, and $n,l$ denote the radial and angular quantum numbers, respectively. To model the transverse part of the wavefunction, $e^{il\theta}R_{n,l}(r)$, we take a circular cross-section with an potential $V = \infty$ for $r > r_0$ and $V = V(r)$ for $r\leq r_0$. We choose potentials of the form studied in ref.~\cite{Tserkovnyak2006}, $V(r) = A(1-(r/r_0)^{b/2})$, where $A = V(0) - V(r_0)$, $V(r_0)$ is the surface potential, and $b \geq 2$ dictates the shape of the potential. This potential is taken to be independent of the number of occupied subbands and to remain constant as the chemical potential in the nanowire is varied. In the results of later sections, $\sim 10$ subbands enter into the description of device conductance. The mobile charge induced in the nanowire when 10 subbands are occupied is an order of magnitude smaller than the total charge corresponding to a typical surface state density of $10^{12}$ cm$^{-2}$\cite{Noguchi1991}, assuming all surface states are ionized. This justifies an approximate treatment of the radial potential as fixed and independent of the carrier density. Since the surface charge density is positive (surface states are donor-like), the conduction band usually bends downward \cite{Noguchi1991}, and in this paper we consider $A \geq 0$. The Hamiltonian for a single conduction electron including an applied axial magnetic field can be written \cite{Tserkovnyak2006}:
\begin{equation}
\begin{split}
&H = \frac{-\hbar^2}{2m^{*}}\left[\frac{1}{r}\frac{\partial}{\partial{r}}+\frac{\partial^2}{\partial{r}^2}+\frac{\partial^2}{\partial{z}^2}\right.
\\
&+\left.\frac{1}{r^2}\frac{\partial^2}{\partial{\theta}^2}-\frac{1}{r^2}\frac{2}{\hbar}\phi_zL_z-\frac{1}{r^2}\phi_z^2\right] + V(r),
\label{eq1}
\end{split}
\end{equation}
where $\phi_z =\Phi/\Phi_0 = \pi{B_z}r^2/(h/e)$ is the normalized magnetic flux, $B_z$ is the axial magnetic field, $L_z$ is the orbital angular momentum operator and $m^{*}=0.023 m_e$ (for InAs) where $m_e$ is the electron mass. Contributions from the Zeeman effect and spin-orbit coupling, which break spin degeneracy and split the subband energies, are neglected (for a more general treatment, see ref.~\cite{Tserkovnyak2006}). For a magnetic field of 8 T (the upper field limit in the experimental section below), the Zeeman energy is $\sim 4.2$ meV for electrons in InAs, smaller than a typical subband energy separation of $10-20$ meV, which justifies an approximate treatment neglecting the Zeeman effect. Equation~\ref{eq1} reduces to the following partial differential equation (PDE) for $R_{n,l}(r)$:
\begin{equation}
E R = \frac{-\hbar^2}{2m^{*}}[\frac{R'}{r}+R''- k^2R-\frac{R}{r^2}(l+\phi_z)^2]+RV(r),
\end{equation}
where primes denote derivatives with respect to $r$, $E=E_{n,l}$ and $R=R_{n,l}(r)$. A 4$^{th}$-order Runge-Kutta PDE solver \cite{Press2007} numerically calculates $R_{n,l}(r)$ at fixed values of $l$ and $\phi_z$. The subband energies $E_{n,l}$ are determined by applying the boundary condition that $R_{n,l}(r_0)=0$. 

\begin{figure}[t]
\includegraphics[width= 8.6cm]{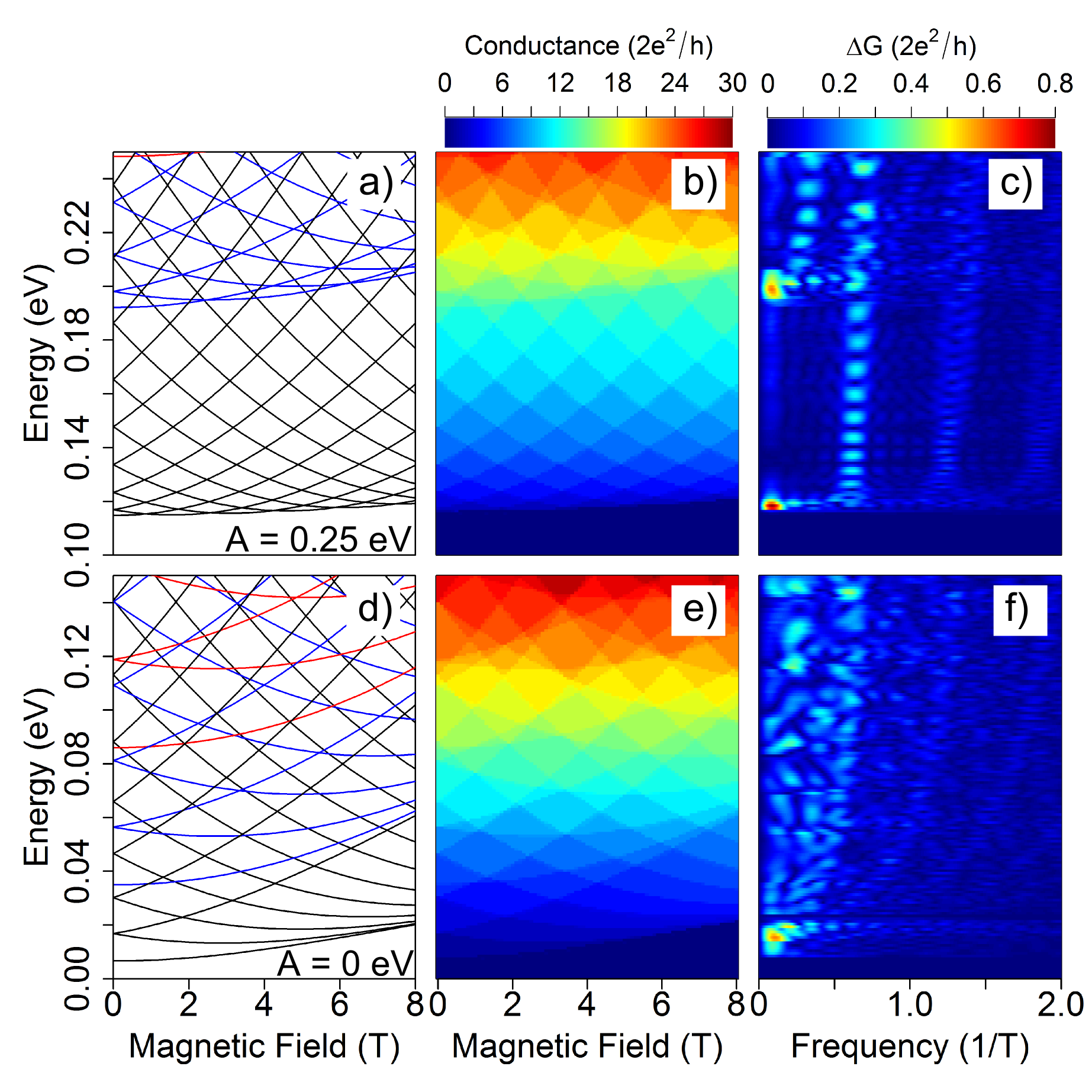} 
\caption{(a,d) Calculated energy levels $E_k(B_z)$ in a nanowire with radius $r_0$ = 38 nm, for radial potentials $V(r)$ with $b$ = 2.75 and $A$ = 0.25 eV (a) and $A$ = 0 eV (d). The radial quantum number is distinguished by color, where black denotes $n=0$, blue $n=1$, and red $n=2$. In (a), the curvatures of $E_k(B_z)$ in the $n=1$ manifold, appearing above 0.19 eV, are smaller than those of the radial ground state manifold because the radial expectation value $r_{\text{eff}}$ is closer to the nanowire center for $n>0$. In the lower part of (d), the successive subband minima move upwards in energy. This is due to an effective increase in confinement as the quantum number $|l|$ increases, since the wavefunction becomes more narrowly peaked. (b,e) Ballistic magnetoconductance calculated from the energies in (a,d) using the Landauer equation. The conductance increases (decreases) stepwise by $2e^2/h$ when a new transverse mode is populated (emptied). The vertical axis is to be identified with the chemical potential in the nanowire, modulated by gate voltage. (c,f) Fast-Fourier transform (FFT) of the conductance in (b,e). The colorscale is labeled $\Delta G$ since the FFT peak intensity reflects the amplitude of magnetoconductance oscillations at a particular frequency. The mean of each conductance trace is subtracted prior to the FFT in order to avoid low frequency artifacts.} \label{fig2}
\end{figure} 

\section{Results}
\subsection{Theory} 
\indent In this section we calculate magnetoconductance for a cylindrical InAs nanowire FET, assumed to be in the ballistic transport regime, in order to establish a qualitative picture for how the radial potential determines the pattern of conductance versus field and gate voltage. Generalization to the diffusive transport regime is discussed in the experimental section. A nanowire radius $r = 38$ nm, similar to experimental value, is chosen for the calculations. Figure 1b shows a schematic of the typical FET geometry; however, in the calculations which follow we assume no breaking of cylindrical symmetry by the back gate, which is approximately justified when the gate oxide thickness is large compared to the nanowire diameter. First is considered the case of strong band bending, taking $A$ = 0.25 eV and $b$ = 2.75. Figure 1c shows that in this case, the electron distribution is mostly independent of $l$, and is concentrated near the nanowire surface, consistent with the expected accumulation layer. In contrast, figure 1d shows that for a flat potential, the $l=0$ and $l\neq 0$ states have very different spatial distributions. The magnetic field dependence of the subband energies is intuitively understood by imagining that electrons are located near the peak of the wavefunction. The limiting case of strong band bending is a two dimensional electron gas (2DEG) near the surface considered in ref.~\cite{Richter2008}, where the subband energy (in the radial ground state, $n=0$) is given by: $E_l =  \frac{\hbar^2k_z^2}{2m^*} + \frac{\hbar^2}{2m^*r^2_{\text{eff}}}(l-\phi_z)^2$, and $r_{\text{eff}}$ is the electron's average radial position. Figure 2a shows that strong band bending in our model also produces $n=0$ energy bands that are nearly parabolic with respect to the magnetic flux. \\
\indent Assuming quasi-ballistic conditions, electrical conductance may be calculated using the Landauer equation \cite{Ferry2009}, $G = 2e^2/h\sum_m\int{\tau_m}(E)(df/dE)dE$, where $\tau_m(E)$ is the transmission probability for the $m^{th}$ subband, and $f=f(E,T)$ is the Fermi-Dirac distribution at temperature $T$. For ballistic transport, $\tau_m(E)$ is a step function of unit height centered at the subband energy $E_m(B_z)$. The resulting conductance in the presence of strong band bending is shown in figure 2b. This gives a series of conductance steps of height $2e^2/h$ occurring when a subband crosses the chemical potential, defined here as $\epsilon = E_F - E_C$, where $E_F$ and $E_C$ are the Fermi energy and conduction band edge, respectively. The rounding of the conductance steps is determined by the temperature in the Fermi-Dirac distribution, which in figure 2 is set to $T = 1$ K.  \\
\indent The frequency components of the magnetoconductance oscillations may be analyzed by calculating the Fourier transform with respect to magnetic field at each value of $\epsilon$, as shown in figure 2c. The mean value of each conductance trace was subtracted prior to performing the fast Fourier transform (FFT) in order to suppress artifacts from the dc component of magnetoconductance. In the region below $0.19$ eV, where only the radial ground state ($n$ = 0) is occupied, the FFT shows a dominant peak at a frequency $\sim 0.65$ T$^{-1}$. This peak occurs when the flux enclosed by the average electronic radius is equal to $\Phi_0$. A frequency of 0.65 T$^{-1}$ implies an effective radius $r_{\text{eff}} = 29$ nm, consistent with radial wavefunctions shown in figure 1c. The slight increase in frequency of this peak as the chemical potential increases is due to the occupation of states with higher angular momentum that have $r_{\text{eff}}$ closer to the nanowire surface. The peaks at double and triple this frequency are harmonics that arise from taking the FFT of a square-like wave, and are unrelated to mesoscopic interference effects. For example, the Al’tshuler-Aronov-Spivak (AAS) effect for cylindrical shell conduction \cite{Altshuler1981,Bachtold99} would produce a peak at twice the fundamental frequency (i.e. corresponding to a flux of $\Phi_0/2$), however this is not included in our model, and we see no evidence for such oscillations in the experiments of the next section. Above $0.19$ eV, an additional peak appears at lower frequency, due to the first radial excited state manifold. The effective electronic radius corresponding to this state encloses a smaller flux, resulting in a lower frequency magnetic oscillation.\\
\indent The effect of decreased band bending is shown in figure 2d, where $A=0$, and larger differences are seen in the curvatures of the energies $E_m(B_z)$ between subbands with the same $n$ but different $l$ values. For $A=0$, the radial wavefunctions at zero magnetic field are Bessel functions of order $l$. The transverse wavefunctions for $l=0$ and $l=4$ in the radial ground state ($n=0$) are shown in figure 1e. For $l= 0$, the radial wavefunction is concentrated in the center of the nanowire, giving a nearly flat magnetic field dependence of the lowest energy level in figure 2d. As $|l|$ is increased, the wavefunction peak moves toward the surface, with successively greater curvature in $E_m$ versus $B_z$. The flat potential also lowers the energies of radial excitations, reordering the filling of states as the chemical potential is increased in comparison to strong band bending. Figure 2f shows the FFT of the magnetoconductance for $A=0$. Rather than distinct peaks, it shows a distribution of frequencies that correspond to a wider distribution of effective electronic radii compared to strong band bending. Generally, the structure of the energy spectrum is not strictly periodic in chemical potential or magnetic field. This ensures that by probing the magnetoconductance over a sufficiently large range of chemical potential (gate voltage), a fingerprint of the radial electrostatic potential can be obtained, in principle. The results in figure 2 make it clear that a flatter potential produces a more aperiodic conductance pattern that would allow the corresponding subbands to be more easily identified by comparing theory to experiment. It also suggests that for a flatter potential and many occupied subbands, the AAS interference effect should be washed out by there being a range of effective electronic radii. To check for self-consistency between theory and experiment, the conversion between gate voltage and chemical potential is straightforward to estimate based on the geometrical capacitance \cite{wunnicke2006} of the gate and the density of carriers in the nanowire. \\
 \begin{figure}[t]
\includegraphics[width= 8.6cm]{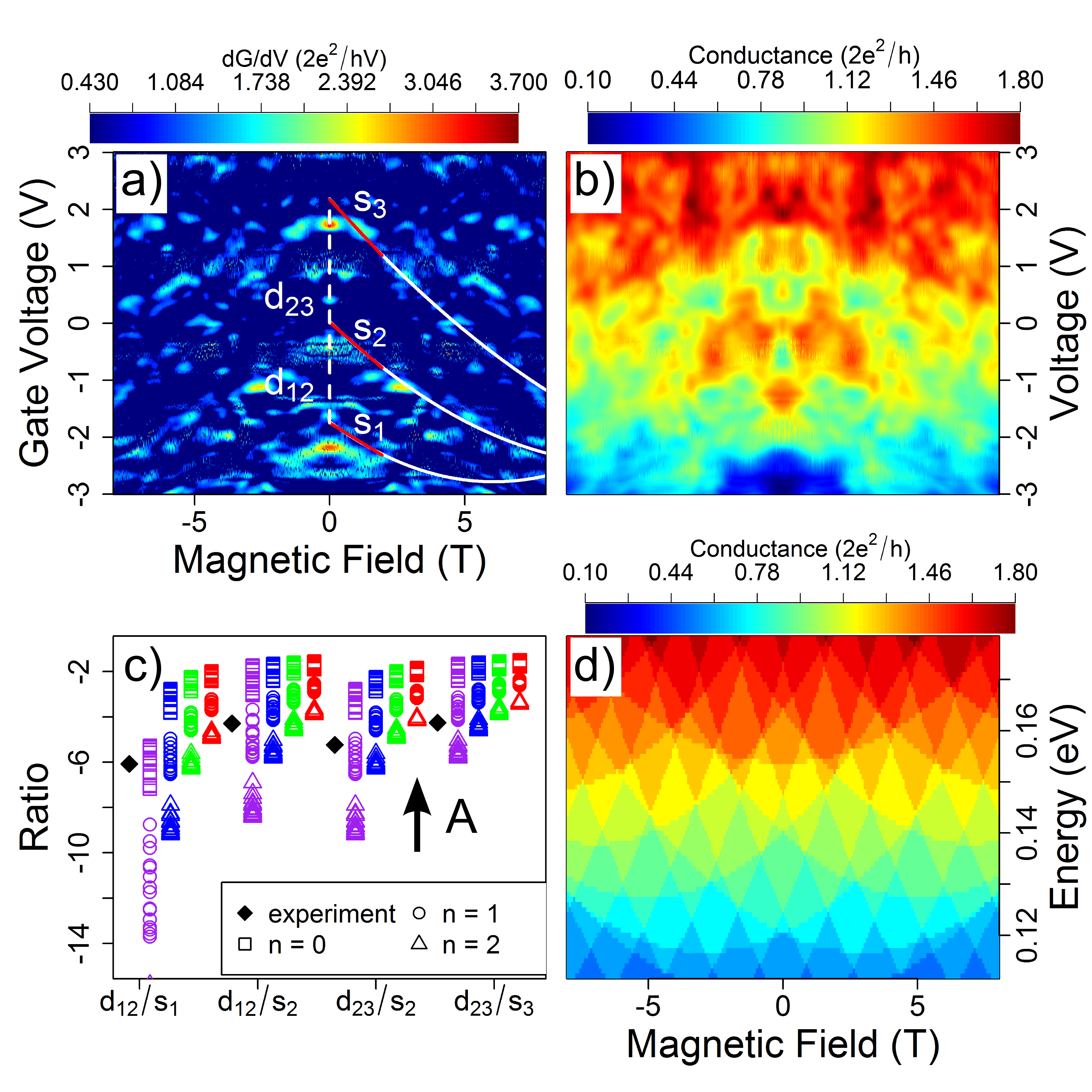} 
\caption{(a) Derivative with respect to gate voltage of the experimental conductance of the InAs nanowire FET, where values below 0.43 $\frac{2e^2}{hV}$ have been removed for clarity. White curves are least squares fits to parabolas consistent with transverse subbands. Red lines are linear fits to the parabolas from 0 to 2 T, used for extracting the zero-field slope of each curve. $s_1 - s_3$ indicate the slopes and $d_{12}, d_{23}$ indicate the vertex separations in gate voltage. (b) Raw experimental magnetconductance for this device, the source of the data shown in (a). (c) Ratios $d_{i, i+1}/s_{j}$ calculated across the parameter ranges $0 \leq A \leq 0.2$ eV and $2\leq b\leq 9$. Colors indicate different sets of $l$ values: purple: $l = (-1,-2,-3)$, blue: $l = (-2,-3,-4)$, green: $l = (-3,-4,-5)$, red: $l = (-4,-5,-6)$. These are plotted for three different radial excitation manifolds, $n=0,1,2$. Experimental values (black diamonds) from the fits to the data in (a) show best overall agreement with the $n=1$, $l = (-1,-2,-3)$ states. The $\uparrow A$ symbol indicates the direction of increasing $A$ values, i.e. stronger band bending. (d) Simulated magnetoconductance for $A=0.11$ eV and $b=2.75$, over a range of chemical potential consistent with the experimental data, as described in the text. The subband transmissions are of order $L_0/L$ (see text), where $L_0 = 20$ nm is a characteristic length on the order of the mean free path, and is a free parameter for matching the simulated conductance to the experimental values. Note that the conductance scales in (b) and (d) are the same. } \label{fig3}
\end{figure} 
\subsection{Experiment}
\indent A FET device based on an InAs/In$_{0.8}$Al$_{0.2}$As core/shell nanowire, with a nominal Te doping density in the shell of $5\times 10^{16}$ cm$^{-3}$, was investigated experimentally. The core radius was estimated from scanning electron microscopy to be $r_0 \approx$ 38 nm, and a channel length $L \approx 200$ nm between contacts was fabricated. A 300 nm thick gate dielectric (SiO$_2$) separated the nanowire from the backgate. The device geometry is shown in figure 1b, and the fabrication procedure was described previously \cite{Gupta2013}. The chemical potential is controlled by modulating the voltage of the backgate, and an axial magnetic field up to 8 T is applied. As mentioned above, a gate separation much larger than the nanowire diameter is crucial to minimize the breaking of cylindrical symmetry in the nanowire radial potential when a gate voltage is applied. Conductance data shown in figures 3a and 3b was measured at a lattice temperature of 30 mK, with an estimated electron temperature $\approx 100$ mK. Similarly, a temperature of 100 mK was used for the simulations shown in figures 3c and 3d. At this temperature, the device conductance typically shows additional modulations with field and gate voltage due to electron interaction effects (Coulomb repulsion) and interference effects (e.g. universal conductance fluctuations), the latter being due to a phase coherence length comparable to the channel length. The details of these effects are not amenable to simulation because they depend on device specific, mesoscopic potential fluctuations, i.e. they are essentially random in nature. Experiments carried out at higher temperatures, such that the the phase coherence length is suppressed but the subband level spacing is still large compared to thermal energy, could suppress some of the conductance modulations seen in figure 3b that are unrelated to the subband effects. On the other hand, we expect a phase coherence length comparable to the nanowire circumference is necessary in order for the theory of the previous section to be applicable. We first note two caveats in comparing the experimental data to the model described previously. One, the experimental magnitude of conductance indicates this device to be in the diffusive, weakly localized regime, rather than the quasi-ballistic regime. This is accounted for in the model by calculating conductance using $G = (2e^2/h)N^2L_0^2/(NL_0L+L^2)$, where $N$ is the number of occupied channels, and $L_0$ is a characteristic length of the order of the mean free path \cite{Datta95}. Note that this expression is derived from the Landauer equation to approximately include the effects of elastic scattering and quantum interference, and is only valid in the weakly localized regime where transport is phase coherent and $NL_0\gtrsim L$. For the case of  $NL_0 \gg L$, this equation simplifies to $G = (2e^2/h)NL_0/L$, which is the Landauer result with all transmission probabilities given by $\tau_m(E) = L_0/L$. We were not able to measure the field effect mobility directly, as the backgate was not sufficient to pinch off the conductance. From measurements of longer channel devices using nanowires of the same growth batch, we find an average elastic mean free path of $\lambda \sim35 \pm 13$ nm. Secondly, the back-gated geometry breaks the cylindrical symmetry of the nanowire and produces, at a finite gate voltage, an asymmetric radial potential. We have not included this effect in the modelling, however a numerical estimate suggests it will not be a dominant effect. Using a finite element model \footnote{Finite element method calculations were performed using COMSOL Multiphysics v4.2a.} of our device geometry, a difference in surface potential between the top and bottom surfaces of the nanowire is found to be $\approx -6.7$ mV per volt of applied gate voltage. At the largest gate voltage, $\pm 3$ V, this yields only $\sim 20\%$ of the typical surface band bending $\approx 100$ meV.  Note the gate sweep is also centered around $V_g = 0$ in order to minimize this effect.\\
\indent As described in the previous theory section, positive steps in conductance occur as the chemical potential is increased at a fixed magnetic field. Hence, the derivative of conductance with respect to gate voltage should give a positive value when the chemical potential is equal to the energy of a transverse subband, and be zero elsewhere. In figure 3a we plot the derivative of the raw conductance data shown in figure 3b. The data shows three plausible parabolic trajectories where the derivative has an average value above the noise floor. We find analytic expressions for these curves by averaging the points in the vicinity of these features and fitting to quadratic functions with least squares fitting. The resulting curves are plotted as the three white lines in figure 3a. The parabolic fit describing one subband does not contain enough information to identify the subband, since it can be reproduced by a variety of radial potentials and $n,l$ values. However, several curves can provide sufficient information to assign the subbands. We construct a simple quantitative measure by defining $d_{i,i+1}$ as the energy separation between adjacent subbands at zero magnetic field, and $s_i$ as the linear slope near zero field (calculated from 0 to 2 T). The ratio $d_{i,i+1}/s_j$ is limited to a certain range of values that depend on the $A$ and $b$ parameters describing the radial potential. Examples calculated from the model are shown in figure 3c for $l$ values from -1 to -6 and in three radial manifolds, $n=0,1,2$. Here $A$ is varied from 0 to 0.2 eV, and $b$ from 2 to 9 (however, the ratios depend much more strongly on $A$ than $b$). The magnitude of $|d_{i,i+1}/s_j|$ decreases as $A$ is increased, i.e. as the surface potential becomes larger. The dependence on $b$ is opposite to this, but much weaker. This provides an unambiguous way to correlate the parabolic features in the experimental data to a model of the radial potential, and in principle to identify the corresponding transverse subbands. Note that the ratio $|d_{i,i+1}/s_j|$ is independent of the energy scale, so that an a priori correspondence between experimental gate voltage and chemical potential is not needed for matching theory to experiment; rather, finding a match using these ratios automatically determines the correspondence. Clearly, a stronger assignment can be made when there are more subbands visible in the data. From the data in figure 3a we extract the ratios indicated by black diamonds in figure 3c. For three out of the four possible ratios, the subbands with $n$ = 1, $l$ = -1 to -3 match the data. The average band bending parameter for these three points is $A=0.11$ eV. We conclude that these states are likely candidates to assign to the three parabolic features, however, the conductance data from this device is too complicated by other effects in the weak localization regime to make an unambiguous assignment. \\
\indent In figure 3d we simulate the conductance for a radial potential with $A$ = 0.11 eV and $b$ = 2.75, which gives a reasonable match to the experimental conductance in figure 3b. This matching suggests that $V_g=0$ V corresponds to a chemical potential of about 140 meV, and the gate range of $\pm 3$ V corresponds to an energy shift of about 70 meV. This is crudely checked by estimating the gate modulation of carrier density via the expression $\Delta n = \frac{C_g}{\mathcal{A}eL}\Delta V_g$, where $n$ is carrier density and $\mathcal{A}$ is the nanowire cross-sectional area. $C_g$ is the geometric gate capacitance which we estimate as $8.6$ aF. The gate range of $6$ V corresponds to $\Delta{n} = 3.55 \times 10^{-17}$ cm$^{-3}$. Alternatively, the dependence of $n$ on chemical potential $\epsilon$ can be calculated in the diffusive regime. Here we use an expression for carrier concentration appropriate to a nanowire with transverse subbands: $n(\epsilon) = \frac{\sqrt{2m^*k_BT}}{\hbar\pi\mathcal{A}}\sum_iF_{-1/2}(\frac{\epsilon-E_i}{k_BT})$ \cite{Gupta2013}, where $k_B$ is the Boltzmann constant, $F_{-1/2}$ is the Fermi Dirac integral of order -1/2, and $E_i$ are the subband energies below $\epsilon$. The range of $\epsilon$ in figure 3d corresponds to $\Delta{n} = 1.28 \times 10^{-17}$ cm$^{-3}$, which gives 0.36 times the value estimated from gate capacitance. However, these quantities are of the same order, and we have not taken into account gate screening in the short channel device that would lower $C_g$ and reduce $\Delta{n}/\Delta V_g$. The experimentally observed change in conductance $\Delta G \approx 1.6 \times 2e^2/h$ over the 6V gate range is consistent with reasonable values for the chemical potential and the average mean free path in a diffusive transport picture. Using $6$V$=\Delta V_g = \frac{L^2}{C_g \mu}\Delta G$, we obtain an effective mobility $\mu = 960$ cm$^2$V$^{-1}$s$^{-1}$, corresponding to a mean free path $\lambda = 18$ nm when setting $\epsilon = 0.14$ eV. Taking into account gate screening by the contacts and/or mobile charges associated with the oxide or interfaces would decrease $C_g$, implying slightly larger values for mobility and mean free path.\\
\indent For transport through many modes in a phase coherent conductor, quantum interference effects can lead to a non-negligible contribution to conductance. If a system is in the weakly localized regime such that $NL_0 \gtrsim L$, the total conductance can be approximated as $G = (2e^2/h)N^2L_0^2/(NL_0L+L^2)$ \cite{Datta95}. This approximation applies to a 1D system with elastic backscattering, where all subband transmission probabilities are equal. The nanowire studied here satisfies the first assumption since it is a quasi-1D system with large enough separations in subband energies to strongly suppress interband scattering. This is confirmed numerically by calculating the transition rates between different subbands caused by a perturbing potential. From previous studies of nanowire conduction we have found that Coulomb scattering due to surface charge defects dominates electron mobility at low temperature \cite{Gupta2013}. Therefore the interband transition rate is calculated using the Coulomb potential of a random assembly of surface charges at a density of 10$^{12}$ cm$^{-2}$. Under these conditions, we find that interband transitions are indeed suppressed by several orders of magnitude compared to back scattering. Additionally, the same calculations show that comparing the back scattering rates of all subbands included in figure 3d at $k=0$ for each subband yields at most a $2\%$ difference. This validates the assumption that all subbands have nearly the same transmission probability. Simulation of planar potential jumps to mimic the effects of stacking faults yields similar results. Thus, although an electron scatters elastically several times while transiting the FET, it is very likely to remain in the same subband, and the subband effects predicted in the ballistic model should remain visible, despite a lower overall magnitude of conductance. Finally, this device satisfies the weak localization criterion that $NL_0 \gtrsim L$ for most of the conductance range, since the simulation shows between 9 and 24 transverse states are occupied and we expect a mean free path of $\sim 18$ nm. Using the weak localization equation for conductance, a very good match to the experimental conductance range is found for $L_0 = 20$ nm. This agrees with the $18$ nm mean free path estimated above, and is close to the lower end of range $\lambda \approx 35 \pm 13$ nm obtained from mobility measurements on other nanowires from the same growth batch.\\
\indent The present model does not include mesoscopic interference effects such as Aharanov-Bohm (AB) and Al’tshuler-Aronov-Spivak (AAS) oscillations \cite{Altshuler1981,Bachtold99} that apply to the case of cylindrical shell conduction. Indeed, a phase coherence length $L_{\phi} \approx 275$ nm $> L$ is estimated for this device based on an analysis of the two-point correlation function of magnetoconductance fluctuations \cite{Blomers2008}. With that technique, we obtained similar values of $L_{\phi}$ for several other FETs fabricated with nanowires from the same batch. While the AB effect is suppressed by disorder, the AAS effect should survive and exhibit conductance oscillations with a period of $\Phi_0/2$. However, these effects are most clear and strong in the limiting case of shell conduction at a fixed radius, where all electronic states enclose the same flux. Our results suggest intermediate band bending in this device and therefore a distribution of effective radii, which is expected to strongly attenuate AAS oscillations. Also, the AAS effect should produce oscillations whose phase is independent of the chemical potential (gate voltage), and no such gate-independent oscillation is visible in the conductance data. \\
\section{Conclusion}
This paper has described a model of magnetoconductance based on the energy spectra of transverse electronic states in a semiconductor nanowire. It extends previous work in this area to examine the contrasting effects of weak and strong surface band bending on the patterns of conductance versus magnetic field and gate voltage. Conductance features from experiments on an InAs nanowire were shown to be consistent with the model, and provide a plausible match to specific subbands, although the assignment for this particular device is not definitive. Although the device is in the weakly localized regime, characterized by several elastic scattering events per transit, back scattering is found to dominate over interband scattering so that the subband effects on magnetoconductance predicted for ballistic transport should still be visible here. We suggest that in a quasi-ballistic nanowire FET, quantitative analysis of  magnetoconductance patterns using the method of determining the $d_{i,i+1}/s_j$ ratios described previously will allow unambiguous identification of the subbands participating in transport. It can also determine, to a degree consistent with the quality of the data, the radial potential $V(r)$. Cleaner transport can be achieved either by using materials with higher mobility, such as InSb \cite{Plissard2012}, using core-shell nanowires with lower defect densities than the one examined here, or by fabricating shorter channels. There are two caveats to further shortening the channel: it will produce quantization of the axial states, which complicates the conductance calculation, and it will increase gate screening by the nanowire contacts, which reduces the effectiveness of the gate in modulating the chemical potential. Including Zeeman and spin-orbit effects in the model is straightforward \cite{Tserkovnyak2006}, and is expected to improve agreement with experiment at high magnetic fields. It will also provide a method for measuring the subband-specific magnitudes of the $g$-factor and the spin-orbit coupling, assuming the subband splittings due to these effects are visible. Inclusion of the potential asymmetry due to a backgate geometry in numerical simulations is also straightforward, although we estimate this asymmetry to make only a small correction to the surface potential when the gate oxide is sufficiently thick. Accurate understanding of the radial potential and subband structure has implications for controlling surface scattering and tuning the number of modes participating in transport, leading to improved engineering of nanowire devices. \\

\textbf{Acknowledgements --} We acknowledge the Canadian Centre for Electron Microscopy, the Centre for Emerging Device Technologies, and the Quantum NanoFab facility for technical support. Shahram Tavakoli provided assistance with MBE and Roberto Romero provided technical assistance. We thank B. Reulet and M. Khoshnegar for helpful discussions. This work was supported by NSERC, the Ontario Ministry for Research and Innovation and the Canada Foundation for Innovation. G. W. H. and K. W. acknowledge support from the Waterloo Institute for Nanotechnology. 

\bibliography{magneto}
\end{document}